\documentclass[12pt]{iopart} \usepackage{amsfonts,amssymb}
\usepackage{iopams,cite} \usepackage{graphicx} \usepackage{color}

\newcommand{\text}[1]{\mbox{\scriptsize{#1}}}

\begin{document}

\title[Recent Advances in Understanding Biopolymer Translocation]
{Through the Eye of the Needle: Recent Advances in Understanding
  Biopolymer Translocation}

\author{Debabrata Panja$^{*,\dagger}$, Gerard
  T. Barkema$^{*,\ddagger}$, Anatoly B. Kolomeisky$^{**}$}
\address{$^*$Institute for Theoretical Physics, Universiteit Utrecht,
  Leuvenlaan 4, 3584 CE Utrecht, The Netherlands\\ $^\dagger$Institute
  for Theoretical Physics, Universiteit van Amsterdam, Science Park
  904, Postbus 94485, 1090 GL Amsterdam, The Netherlands\\
  $\ddagger$Instituut-Lorentz, Universiteit Leiden, Niels Bohrweg 2,
  2333 CA Leiden, The Netherlands\\ $^{**}$Rice University, Department
  of Chemistry, 6100 Main Street, Houston, TX 77005-1892, USA}
\eads{\mailto{D.Panja@uu.nl}}

\begin{abstract}
  In recent years polymer translocation, i.e., transport of polymeric
  molecules through nanometer-sized pores and channels embedded in
  membranes, has witnessed strong advances. It is now possible to
  observe single-molecule polymer dynamics during the motion through
  channels with unprecedented spatial and temporal resolution. These
  striking experimental studies have stimulated many theoretical
  developments. In this short theory-experiment review, we discuss
  recent progress in this field with a strong focus on non-equilibrium
  aspects of polymer dynamics during the translocation process.
\end{abstract}


\maketitle

\section{Introduction\label{sec1}}

Translocation, commonly understood to be the transport of polymeric
molecules through nanometer-sized pores and channels (nanopores and
nanochannels in short) embedded in membranes, and related dynamic
phenomena in confined geometries are of fundamental and critical
importance for many processes in chemistry, physics and biology, as
well as for many industrial and technological applications
\cite{lodish,muthu_book,zwolak08,meller03,aksimentiev11,keyser11}.  In
the last two decades significant progress has been achieved in
experimental studies of the translocation process at the
single-molecule level
\cite{muthu_book,meller03,keyser11,dekker07,wanunu08,maglia08,liu10,cockroft08,lieberman10,newton06}.
This has opened up the opportunity for the development of nanopore
devices as a new class of chemical and biological
sensors. Experimental successes have naturally stimulated significant
theoretical efforts to understand mechanisms of polymer dynamics in
nanopores and nanochannels
\cite{muthu_book,zwolak08,mathe05,matysiak06,muthukumar03,wong08},
although many questions remain unanswered.
\begin{figure}[h]
\begin{center}
\includegraphics[width=0.5\linewidth]{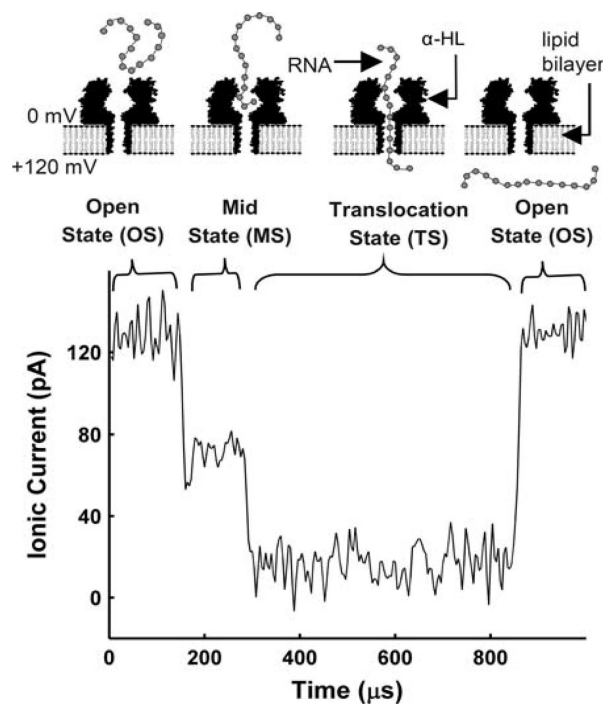}
\end{center}
\caption{A schematic representation of a translocating RNA through an
  $\alpha$-hemolysin pore embedded on a lipid bilayer membrane,
  showing that the blockade of the pore by the polymer coincides with
  the blockade of the ionic current. Since the $\alpha$-hemolysin pore
  cross-section is not uniform, the current-blockade characteristics
  depends on the details of the pore-blockade. Figure reproduced from
  Ref. \protect\cite{butler06}, with permission from Elsevier Inc.}
\label{transfig}
\end{figure}

{\it In vivo\/}, translocation of biological molecules is assisted by
interactions with cellular membranes and/or special protein molecules
\cite{lodish}. {\it In vitro\/}, (bio)polymers are driven across pores
or channels by applying external fields in single-molecule experiments
\cite{muthu_book,zwolak08,keyser11,meller03,dekker07,wanunu08,maglia08,liu10,cockroft08}.
The pores or channels connect two compartments that are separated by
an otherwise impenetrable membrane. A voltage difference is applied
between the two compartments, which would normally cause the flow of
an ionic current through the pore.  When a polymer enters the pore, it
partly blocks the path of the ions, resulting in a ``current
blockade'', a significant decrease in the ionic current. The polymer
makes many attempts to cross the pore; the unsuccessful attempts
result in current blockades of very short durations, which are
digitally filtered out. In the experiments the actual translocation
events are associated with the longer blocked current signals
(Fig. \ref{transfig}). The signature of the ionic current is therefore
of paramount importance in the experiments, as it carries the
signature of the polymeric molecule passing through the pore.

The full details of the current blockade phenomenon are extremely
complex: the involved variables are the pore size, pore geometry,
chemical associations and charge condensations on the pore, ionic
conditions on both sides of the separating membrane (including the
ionic cloud that may condense around the pore), concentration of the
macromolecules in the solution, ionic condensation on the
macromolecules, voltage difference across the pore and temperature.
Some of these variables may not stay constant throughout the typical
duration of an experiment
\cite{muthu_book,zwolak08,aksimentiev11}. Taking all these into
account in detail (without some level of coarse-graining) to model
translocation is beyond the present day capabilities of theory and
computer simulations.

At a broader and more phenomenological level of classification for the
above complexities, the translocation process can be viewed as
controlled by four main factors: (i) external driving fields, (ii)
polymer dynamics, (iii) properties of pores/channels, and (iv)
polymer-pore interactions. In this review we provide a combined
theory-experiment progress report on the understanding of (bio)polymer
translocation. To this end we note that several review articles have
already been published
\cite{muthu_book,meller03,keyser11,milchev,deamer00,deamer02,dekker07,muthukumar07,branton08,movileanu08,makarov08,movileanu09,majd10,gu12}. One
group of reviews
\cite{meller03,deamer00,deamer02,branton08,movileanu08,movileanu09}
tilts heavily towards the experimental side and mostly biological
nanopores are discussed, the second group
\cite{keyser11,dekker07,majd10,gu12} focuses more on artificial
nanopores as biosensing devices, the recent theoretical review 
\cite{milchev} presents a more general theme of polymer dynamics
within confinements with emphasis on computer simulations, while the
book \cite{muthu_book} concerns mostly with theoretical
descriptions of the translocation process from a quasi-equilibrium and
phenomenological angle. An opening therefore remains for reviewing the
developments on (the non-equilibrium aspects of) polymer dynamics of
translocation purely from the theoretical side, relatively unrelated
to the experimental nuances.  That is the main focus of this
review. We do not treat the review as an encyclopedic narrative,
listing the papers and then summarizing their results. Instead, we aim
to provide a unified picture to the reader --- how we perceive the
different studies fitting together (or not): this means that we are
forced to leave out papers that marginally contribute to this unified
picture.

We begin by a glossary in Sec. \ref{secgloss}. Thereafter, in order to
remind the reader of the complexities associated with polymer
translocation, we begin by a concise summary of the experimental
developments in Sec. \ref{sec2}. We then move on to a brief
description of the main conceptual aspects of the translocation
process in Sec. \ref{sec3}. In Sec. \ref{sec4} we compare the
different approaches on polymer dynamics of translocation. In
Sec. \ref{sec5} we summarize the experimental aspects still in
want of theoretical understanding. We finally end this review in
Sec. \ref{sec6} with a brief discussion on where this field is likely
to head to in the coming years.

\section{Glossary\label{secgloss}}

\begin{itemize}
\item[$\bullet$] {\it Persistence length:\/} A mechanical property of
  a polymer quantifying the distance over which it preserves its
  spatial orientation, often denoted by the symbol $l_p$. Molecules
  much shorter than the persistence length resemble straight rods. The
  persistence length of long double-stranded DNA is typically $\sim50$
  nm ($~150$ bp).

\item[$\bullet$] {\it Pore or a channel?:\/} A pore refers to the case
  when the membrane is thin, while a channel refers to the case of a
  thick membrane (i.e., a long pore). The pore or a channel distinction
  depends on the comparison of membrane thickness $t$ with persistence
  length $l_p$ and aperture $a$. For $t \lesssim l_p$, or $t \lesssim a$,
  or both, it is translocation through a pore; otherwise if $t\gg l_p$
  and $t\gg a$ it is translocation through a channel.  Unless otherwise
  stated, this review will consider translocation through a pore.

\item[$\bullet$] {\it Debye length:} It is the length scale by which
the mobile ions screen out the electric field. Its value depends on the
electrolyte concentration used in the experiments, and rarely exceeds a few
nanometers at typical experimental high-salt conditions. In addition,
it is assumed that the electric field that typically drives translocation,
is, for all practical purposes, confined only within the pore.

\item[$\bullet$] {\it A translocation event, dwell time and translocation
time:\/}
  A polymer makes many attempts to cross the pore; the unsuccessful
  attempts result in ionic current blockades of very short
  durations. Although during these times the pore is blocked by
  the polymer, these short blockages are not interesting, and are
  therefore digitally filtered out in experiments. A translocation
  event therefore coincides with pore-blockade across which the polymer
  crosses the pore. The compartment in which the polymer was before
  translocation (resp. in which the polymer is after translocation)
  is called {\it cis\/} --- Latin for `this' (resp. {\it trans\/} ---
  Latin for `across'). The time taken during a translocation event is
  synonymously defined in the literature as pore- or current-blockade
  time, and also known as a dwell time. One has to distinguish the dwell
  time and a translocation time (also known as a transit time) which is
  the average time for a polymer to navigate across the membrane.

\item[$\bullet$] {\it Phantom and self-avoiding polymers:\/} A phantom
  (resp. self-avoiding) polymer is (resp. not) allowed to intersect
  itself. One result that will be used often in this review is that
  the radius of gyration $R_g$ for a self-avoiding polymer of length
  $N$ scales as $N^\nu$, where $\nu$ is the Flory exponent; $\nu=3/4$
  and $\approx0.588$ in two and in three dimensions respectively. The
  radius of gyration for a phantom polymer of length $N$ scales as
  $\sqrt N$.

\item[$\bullet$] {\it Rouse and Zimm polymers:\/} Rouse proposed a
  model for polymer dynamics in 1953, in which hydrodynamic
  interactions among the monomers are completely screened
  \cite{rouse53}; it is known as the Rouse model.  A key
  characteristic of this model is that the equilibration time for a
  polymer of length $N$ scales as $N^{1+2\nu}$, where $\nu$ is the
  Flory exponent. In 1956, the model was extended by Zimm to include
  hydrodynamic interactions among the monomers \cite{zimm56}, the
  corresponding model is known as the Zimm model. The equilibration
  time for a polymer of length $N$ in the Zimm model scales as
  $N^{3\nu}$.

\item[$\bullet$] {\it Anomalous dynamics:\/} The dynamics of a
  particle is called anomalous if its mean-square displacement
  $\langle\Delta s^2(t)\rangle$ in time $t$ scales as $t^\beta$ for
  some $\beta\neq1$; the case $\beta=1$ denotes the ``normal'' or
  Fickian diffusion.

\end{itemize}

\section{A summary of the key experimental developments\label{sec2}}

\subsection{External Fields\label{sec2a}}

It is entropically unfavorable for a polymer molecule to enter into
the pore due to a significant decrease in number of degrees of
freedom, leading to an entropic barrier \cite{muthu_book} (in
Sec. \ref{sec3} we will address this in detail). This entropic barrier
is typically overcome by utilizing external electric fields since most
macromolecules used in nanopore experiments are charged
\cite{henri00,meller02,brun08}. Experiments indicate that increasing
the strength of the electric field decreases the translocation times
exponentially. The external field can also be used to slow down the
threading motion of the polymer molecules \cite{fologea05}.

\subsection{Polymer Dynamics aspects\label{sec2b}}

Originally, translocation experiments have involved only RNA and
single-stranded DNA molecules moving through $\alpha$-hemolysin
biological pores
\cite{kasi96,akeson99,meller00,meller01,howorka01,deamer02,sauer03,
  mathe04,wang04,butler06,dudko07,butler07,wanunu08,lin10}.  These
molecules are very flexible and highly charged, which allows them to
be driven through the pore by an applied electric field. The
translocation events are associated with transient dips in ionic
currents, and the length of these blockades is related to polymer
lengths \cite{kasi96,meller01,meller03}. Analysis of blockade duration
times and currents has indicated that nanopores can successfully
discriminate between different types of polynucleotides
\cite{akeson99,meller00,deamer02}, although a single-nucleotide
resolution has not been achieved, mostly due to polymer fluctuations
\cite{deamer02}. It has been shown also that nanopore translocation
measurements might be used to evaluate the phosphorylation state,
chemical heterogeneity as well as the orientation of entering nucleic
acid molecules with a high sensitivity
\cite{wang04,butler06,butler07,wanunu08}. Striking experiments from
Meller's group \cite{wanunu08} pointed out that it is possible to
distinguish $3'$ or $5'$ end translocations of identical DNA
molecules. This is because different packing and orientation of
individual DNA bases in the channel produce different effective
interactions with the pore, leading to different dynamics that can be
observed in nanopore translocation experiments.

The high sensitivity of nanopore translocations has been utilized
later in creating a single-molecule method for analyzing the dynamics
of processes associated or coupled to DNA and RNA molecules such as
unzipping kinetics of double-stranded DNA molecules and hairpins
\cite{howorka01,sauer03,dudko07,verc01}, DNA-protein interactions
\cite{hornblower07,goodrich07}, helix-coil transitions \cite{lin10}
and processive replication of DNA by polymerase enzymes
\cite{cockroft08,lieberman10}. These experiments have also led to a
development of a new single-molecule dynamic force-spectroscopy method
\cite{dudko07,hornblower07,tabard-cossa09,tropini07}, similar to
existing AFM methods, although with much higher resolution,
sensitivity and robustness that has turned out to be extremely
important for investigations of various biological systems.

The success of translocation experiments for studying DNA and RNA
molecules and related processes have stimulated significant efforts to
utilize this approach to investigate other biological
\cite{movi05,sutherland04,ouk07,wolfe07,mohammad08,bikwemu10} and
synthetic polymers
\cite{bezrukov96,movileanu01,movileanu03,murphy07,ouk08}. Protein
translocations are critically important for successful functioning of
all biological systems since more than 50$\%$ of proteins produced in
cells must traverse cellular membranes \cite{lodish, muthu_book}.
Experimental measurements of transport of polypeptides molecules via
$\alpha$-hemolysin channels indicate that the overall translocation
process can be described by a simplified two-barrier single-well free
energy profile that strongly depends on the strength of the external
electric field and on the length of peptide molecules
\cite{movi05,wolfe07,mohammad08,bikwemu10}. Nanopores have also been
used to analyze the structure of peptide molecules in the case of
collagen related systems where some intermediate conformations have
been observed \cite{sutherland04}. In addition, the nanopore recording
technique was useful for studying protein folding dynamics with a good
sensitivity and a controlled spatial resolution \cite{ouk07}. For
non-biological polymers several studies concentrated on the use of
flexible polyethylene glycol molecules
\cite{bezrukov96,movileanu01,movileanu03} for understanding polymer
partitioning in nanopores. Although these experimental results
suggested that the partitioning follows a simple scaling law of de
Gennes, other experiments \cite{bezrukov96} and theory
\cite{sakaue06,sakaue07a} suggest that there ought to be deviations
from this scaling law. In another study \cite{ouk08}, the
translocation of dextran sulfate molecules has been utilized for
investigating the effect of screening in the transport of
polylelectrolyte molecules. Additionally, the nanopore threading of
another synthetic polyelectrolyte, sodium poly(styrene sulfonate), has
been analyzed as a new way of controlling the transport of
macromolecules for future nanotechnological applications
\cite{murphy07}.

\subsection{Properties of Nanopores\label{sec2c}}

The central part of all translocation processes is a pore that
provides a confined space for polymer motion. Physical and chemical
properties of channels play a critical role in the success of nanopore
experiments \cite{muthu_book}. There are two types of nanopore devices
currently used in studies of polymer transport across the
channels. One of them is based on the biological toxin protein
$\alpha$-hemolysin that inserts spontaneously into membranes, forming
roughly cylindrical pores with a diameter of $\sim 1.5$ nm in the
narrowest part
\cite{meller03,lieberman10,kasi96,akeson99,meller00,meller01,howorka01,deamer02,sauer03,mathe04,wang04,butler06,dudko07,butler07,wanunu08,bezrukov96,movileanu01,movileanu03,murphy07,ouk08,movi05,wolfe07,mohammad08,bikwemu10,an12}.
This channel is always in the open configuration, allowing small
molecules, ions and polymers to go through the membrane. The advantage
of using this biological pore is the fact that it can be chemically
modified via mutations to study different aspects of polymer
translocation
\cite{muthu_book,meller03,movi05,wolfe07,mohammad08,bikwemu10}. In
experiments performed by Movileanu and coworkers
\cite{movi05,wolfe07,mohammad08,bikwemu10} it has been shown that
mutations introducing negatively charged acidic binding sites at
special positions of the nanopore might significantly facilitate the
transport of cationic polypeptides. In addition, utilization of this
protein channel is advantageous for studying biologically oriented
problems of polymer translocation. The biological nature of the
$\alpha$-hemolysin channel has been successfully utilized for the
observation and explicit measurement of helix-coil transitions in some
polynucleotide molecules \cite{lin10}. However, there are also many
problems associated with applications of biological nanopores, such as
restricted sizes and limited stability with respect to the changes in
external physical and chemical parameters \cite{dekker07,heng04}. It
is worthwhile to note that recently other membrane proteins have also
been utilized as channels for polymer translocation
\cite{cheneke12,mohammad12,manrao12,pavlenok12}.

Stimulated by shortcomings of biological channels, another approach
that utilizes artificial nanopores in solid-state membranes has been
proposed \cite{keyser11,dekker07,wanunu08,cockroft08,liu10,storm03,storm05a,storm05b,heng04,heng05,keyser06,heng06,trepagnier07,mcnally08,smeets08,lu11,li01,li03,stein02, merchant10,wanunu10,niedzwiecki10,niedzwiecki13}.
Solid-state nanopores provide a controllable and reproducible method
of investigation of polymer threading at different conditions that can
be also easily connected with other single-molecule methods
\cite{dekker07,keyser06,trepagnier07,wanunu10}. Polymer translocations
through artificial nanopores have been successfully coupled with
optical-trap devices that allowed to measure explicitly forces that
are driving charged macromolecules through confined regions
\cite{keyser06,trepagnier07}. It has also been observed that the
polymer threading through solid-state channels made from silicon
nitride and related materials might differ from the dynamics observed in
biological channels \cite{storm05b}. These experiments show that
translocation times of a double-stranded DNA through $ \sim10$ nm
pores have a power-law dependence as a function of the DNA length, in
contrast to a linear dependence observed in the $\alpha$-hemolysin
channel. In another set of experiments on solid-state artificial
nanopores, it has been illustrated that double-stranded DNA molecules
can pass the channel in many conformations including linear and
folded states \cite{storm05a}. Recently it has been shown that
solid-state nanopores can be fabricated from non-silicon materials.
Several experimental studies have indicated that single-walled
carbon nanotubes can serve as channels for translocation of
single-stranded DNA molecules \cite{liu10,lee10}. In addition, a new
exciting possibility for the pore transport came with a discovery
of graphene: it was suggested recently that graphene nanopores can
be viewed as perfect ultra-thin pores for polymer translocation
\cite{russo12,garaj10,schneider10,merchant10,venkatesan12}. Another
interesting direction for new nanopore devices has been proposed
recently with the development of opal films \cite{cichelli06}. However,
although the application of artificial channels had many successes
in various scientific fields, synthesized pores also have several
serious disadvantages due to the inability to create reliable
nanopores with small diameters, and complex chemical interactions
with translocating polymers created by the fabrication procedures
\cite{wanunu08,wanunu10,smeets08}. It is important to take into account
different properties of biological and artificial nanopores in order to
understand fundamental properties of polymer translocation.

\subsection{Polymer-Pore Interactions\label{sec2d}}

Nanopore experiments directly measure the pore-blockade time, which is
strongly affected by these interactions
\cite{muthu_book,mathe05,wanunu08,mirsaidov09}. Mirsaidov and
coworkers have shown that it is possible to discriminate DNA with and
without covalent methylation modifications of cytosines
\cite{mirsaidov09}. Using a synthetic biconical nanopore the
permeability of different DNA molecules has been measured with a high
precision, suggesting the critical role of polymer-pore interactions.
Using tethered oligonucleotides Howorka and Bayley \cite{howorka02}
have determined the electric potential within protein pores,
supporting the idea that the main potential drop is taking place
across the $\beta$-barrel of the $\alpha$-hemolysin channel. The
sensitivity of nanopore experimental methods for polymer-pore
interactions have been also demonstrated in striking experiments on
single-stranded $3'$ and $5'$ end DNA translocations
\cite{mathe05}. Based on pore-polymer interactions, nanopore
techniques as a new mass-spectroscopic method for separation of
macromolecules has also been proposed and successfully utilized \cite{robertson07,reiner10,baaken11}.

\section{Generic description of the translocation process\label{sec3}}

\subsection{Translocation is an activated process\label{sec3a}}

A key generic property revealed by many of the experiments is that
translocation is an activated process. The activation barrier is of
entropic origin: the polymer enjoys much more configurational
possibilities in the bulk --- far away from the membrane (where the
pore/channel is embedded in) --- than when it is threaded through the
pore. The height of the barrier, therefore, can be theoretically
estimated as follows. For a self-avoiding polymer of length $N$, the
number of configurational states per volume, accessible to it in the
bulk scales as $Z_b(N)\approx A\mu^N N^{\gamma-1}$ in which $\gamma$
is a universal exponent --- $\gamma=49/32$ and $\gamma\approx1.16$ in
two and three dimensions respectively --- while $A$ and $\mu$ are not
universal \cite{diehla98}. The corresponding number of states per
volume for the same polymer, but whose one end has just about reached
the pore, and therefore can be thought of as tethered to the membrane,
is approximated by $Z_w(N)\approx A_1\mu^NN^{\gamma_1-1}$ in which the
parameter $\mu$ is not affected by the introduction of the membrane,
$\gamma_1$ is a different universal exponent --- $\gamma_1=61/64$ and
$\gamma_1\approx0.68$ in two and three dimensions, respectively ---
while $A_1$ is again not universal \cite{diehla98}. Now consider the
translocating polymer, for which there are $n$ monomers on one side
and $N - n$ monomers on the other (assuming that the nanopore is
ultra-thin so that there are no monomers inside the channel). Since
this situation can be seen as two strands of polymers with one end (of
each strand) tethered on the membrane, the number of states for this
polymer is given by $Z_w(n)Z_w(N-n)$, which attains a minimum when
$n=N/2$. The entropic barrier faced by a translocating polymer is thus
\begin{eqnarray} 
\Delta S=\log\frac{Z_b(N)}{Z^2_w(N/2)}=c\log
  N+k, 
\label{e0}
\end{eqnarray} 
with $c=\gamma-2\gamma_1+1$ and $k=\log A - 2\log
A_1+2(\gamma_1-1)\log 2$.

\subsection{The three stages of translocation dynamics\label{sec3b}}

\begin{figure}[h]
\begin{center}
\includegraphics[width=0.8\linewidth]{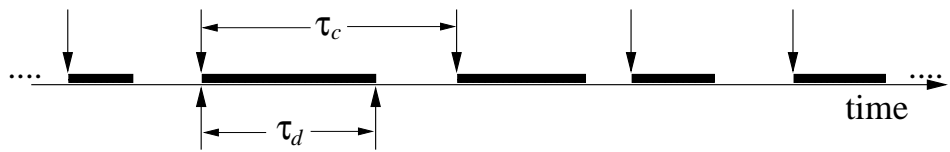}
\end{center}
\caption{A schematic representation of the translocation process,
  where only the pore blockades during translocation events are shown
  by red bars. The captures are indicated by black downward arrow. See
  text for the definition of capture. Relative magnitudes of
  pore-blockade time $\tau_d$ and capture time $\tau_c$ has been made
  out of scale in order to provide visual clarity. The figure provides
  an abstract version of Fig. \ref{transfig}.}
\label{fig1}
\end{figure}

The entropic barrier, obtained above from the partition function is an
equilibrium property. It states that when an ensemble of polymers are
placed in two chambers A and B that are separated by a membrane and
are connected only by a narrow pore, the ratio of the probabilities of
finding a polymer of length $N$ far away from the membrane and being
threaded through the pore with $n$ monomers in one chamber and $(N-n)$
in the other is given by $Z_b(N)/[Z_w(n)Z_w(N-n)]$. Translated to the
case of a single translocating polymer in chamber A, and not an
ensemble of them, it means that the polymer, on its way to chamber B,
will turn back many times to chamber A before it actually succeeds in
translocating. This divides translocation dynamics into three distinct
stages that take place in succession: (i) approach of the polymer in
the vicinity of the pore, followed by repeated threading and
unthreading of one of its ends into the pore, (ii) a final threading
into the pore, which is often referred to as ``capture'' in the
experimental literature, and (iii) the eventual translocation
event. Every time there is a threading event, the polymer occupies a
substantial cross-section of the pore, blocking the ionic current,
although only for short durations. A translocation event corresponds
to the blockade of the ionic current as well, but for longer
times. The long(er) ionic current blockade is therefore preceded by
many short(er) spikes of ionic current blockade events caused by
repeated threading and unthreading events, and are digitally filtered
out in experiments. A sequence of such filtered out current blockade
events is schematically shown in Fig. \ref{fig1}. When the sequence
of current blockade events are followed over a long time, one can
obtain sufficient statistics of the process, from which an average
capture time $\tau_c$ [or equivalently the capture rate
$\tau_c^{-1}$], and the average pore-blockade time $\tau_d$ can be
obtained.

Below we briefly describe the main issues related to the capture
process.

\subsection{The capture process\label{sec3c}}

\subsubsection{Dependence of the capture rate on macromolecular
  concentration $c$\label{sec3c1}}

At the low density of macromolecules used in experiments, the rate
limiting step in the translocation is the availability of the
macromolecules in the immediate vicinity of the pore. Given that the
macromolecules disappear from the {\it cis}-side of the pore, one
should think of the pore as a sink for the macromolecules. Although
there is a voltage across the pore, under typical experimental
conditions the Debye length barely exceeds a nanometer, and the effect
of the field is highly localized in the pore. In such a situation, a
steady state in the macromolecular concentration profile exists on the
{\it cis}-side: at the pore, because of the presence of the sink the
macromolecular concentration is lower, while far away from the pore
the number density is given by $c$. This sets a gradient of
macromolecular concentration on the {\it cis}-side, and generates a
drift of macromolecules towards the sink. The macromolecular current
density $\vec J$ must be proportional to the concentration gradient.

Since the density of macromolecules in experiments is low, these hardly
encounter each other. Thus, all processes, in particular the capture
and translocation processes, will occur with rates that scale linearly
with the overall concentration $c$: if there are twice as many macromolecules,
the total concentration profile will simply scale up by a factor 2, 
and as a result twice as many macromolecules will reach the pore and
succeed in translocation.

\subsubsection{Dependence of the capture rate on other
  parameters \label{sec3c2}}

Given that the capture rate $\tau_c^{-1}$ is also the throughput rate
for translocation, several experiments have studied the capture
phenomenon using both biological pores such as $\alpha$-hemolysin as
well as synthetic pores
\cite{henri00,meller02,chen04,ouk07,wanunu08,wanunu10}. These reveal
four key characteristic properties of the capture rate:
\begin{itemize} 
\item[(i)] It is proportional to the macromolecular volume
  concentration in the buffer solution
  \cite{henri00,meller02,chen04,ouk07,wanunu08,wanunu10}. Typical
  concentrations of macromolecules used in these experiments are a few
  $\mu$g/ml. At these concentrations molecules do not interact with
  each other, and the rate-limiting step for the throughput of a
  translocation experiment is the capture process, as $\tau_d\ll
  \tau_c$.
\item[(ii)] It depends exponentially on the bias voltage $V$ applied
  across the pore above a threshold value; i.e., there is an
  activation barrier for the capture process
  \cite{meller02,ouk07,wanunu08,wanunu10}. E.g., in Ref.
  \cite{meller02} poly(dC)$_{40}$ was translocated through
  $\alpha$-hemolysin pores in a 1M KCl buffer, and data for $\tau_c$
  vs. $V$ demonstrate the existence of two different activation
  barriers at low and high values of $V$, with a sharp crossover at
  approximately 130 mV. In a more recent study \cite{wanunu10}, for
  the translocation of $\lambda$-phage DNA through synthetic SiN
  nanopores in the same buffer a single exponential was reported for
  smaller molecules, but for longer DNA molecules the relation between
  the capture rate and $V$ was found to be linear (although the data
  for longer DNA molecules can also be fitted with an exponential).
\item[(iii)] The length dependence of the capture rate is far less
  clear. Stretched exponential behavior of the capture rate up to a
  certain length of the polymer (and length-independent behavior
  thereafter) has been reported in Ref. \cite{wanunu10}, presently
  there is no theoretical understanding for this behavior.
\item[(iv)] The capture rate is dramatically influenced by the
  application of a salt gradient across the pore: a higher
  (resp. lower) salt concentration on the {\it trans}-side enhances
  (resp. reduces) the capture rate. E.g., in the experiments of
  Ref. \cite{wanunu10}, a 20-fold increase in the KCl concentration on
  the {\it trans}-side of the membrane increased the capture rate by
  almost a factor 30. Although this enhancement was originally thought
  to be due to electroosmotic flow, it has recently been established
  that the enhancement is caused by pure osmotic flow (of water) from
  the low salt concentration side to the high salt concentration side
  \cite{hatlo11}. The DNA is simply dragged along the water flow, much
  like logs floating along a stream.
\item[(v)] Another surprising feature of the capture process reported
  in Ref.  \cite{chen04} is the observation of a capture radius
  $\sim3\,\mu$m around the pore. The experiments are performed with
  synthetic silicon nitride nanopores and $16.5\,\mu$m long
  $\lambda$-DNA. The existence of the capture radius around the pore
  --- which means that once a DNA molecule entered this volume, it did
  not escape during the time of experiment --- was observed using
  fluorescence spectroscopy. The electroosmotic flow has been posited
  to explain this phenomenon \cite{muthu07}.

  For this explanation however, the macromolecule was considered to be
  a point mass located at its center-of-mass, and therefore was not
  considered to be an extended object. Indeed, if the DNA is
  considered to be an extended object, then a back-of-the-envelope
  calculation --- even with the assumption that it is not stretched or
  deformed close to the pore at the micron scale --- it is possible to
  argue that the capture radius should be in the range of 2 microns,
  and the fluorescent spectroscopy images are not precise enough to be
  able to resolve differences between 2 and 3 microns. E.g., at $16.5$
  $\mu$m length, and with persistence length $l_p=33$ nm, the DNA in
  the experiment \cite{chen04} consisted of $N=500$ persistence length
  segments. For a polymer chain, whose conformation obeys Gaussian
  statistics the end-to-end length is $\sim l_p\sqrt{2N}\approx1$
  $\mu$m. However, the DNA is not described by Gaussian statistics,
  but by self-avoiding polymer statistics; assuming that the
  conversion prefactor between Gaussian and self-avoiding prefactor is
  $\approx1$, the end-to-end length of the DNA is $\sim
  l_p(2N)^\nu\approx1.8$ $\mu$m, where $\nu$ is the Flory
  exponent of the polymer $\approx0.588$ in three dimensions. Further,
  since we expect the capture process to be dominated by fluctuations
  (i.e., the polymer finds the pore by fluctuations), one needs to
  consider the statistics of the furthest points of a polymer chain,
  and not simply the end-to-end extent in space. This brings another
  factor $1.2$, implying that one can expect the capture to take place
  when the macromolecule's ends are about
  $1.2l_pN^\nu/\sqrt{2N}\approx2.2$ $\mu$m apart from each other. This
  number is close enough to the value $\sim 3$ $\mu$m for the capture
  radius, observed through fluorescence spectroscopy.

  One implication of this alternative explanation is that one would
  observe a higher capture radius for longer macromolecules. However,
  at the time of writing this review, we are not aware of any
  systematic study on the capture radius as a function of
  macromolecule length.
\end{itemize} 

\subsection{Entropic barrier and pore-blockade time for long
  polymers\label{3d}}

As already discussed in Sec. \ref{sec3a}, at a single-polymer level
the activation barrier manifests itself in repeated
threading-unthreading events before the polymer actually manages to
translocate. {\it The pore-blockade time, which is obtained after
these repeated threading-unthreading events are filtered out, is
in fact independent of the entropic barrier\/} \cite{wolt06}. The
pore blockade events are controlled by polymer dynamics, polymer-pore
interactions, properties of nanopores and external fields. Of these,
the last three have been well-covered in a recent book
\cite{muthu_book}; we therefore take up the issue of polymer dynamics
and the pore blockade time in detail in the following section.

\section{Polymer dynamics and pore blockade time\label{sec4}}

There exists a substantial body of literature, mostly due to theorists
and simulators, who have been interested in translocation as a peculiar
example of a wide family of related activated processes in statistical
physics, including for instance also nucleation theory.  These studies
are therefore largely disconnected from experimental considerations,
unless they address generic aspects of translocation as an activated
process.  In this section we divide them in three categories: (i)
unbiased translocation (translocation in the absence of external driving,
i.e., purely by thermal fluctuations) (ii) field-driven translocation
(translocation driven by a field essentially acting at the pore ---
the field can be of different origin, such as a physical electric field,
an entropic force or a chemical potential gradient), (iii) translocation
mediated by pulling the lead monomer by optical tweezers.
\begin{figure}[h]
\begin{center}
\begin{minipage}{0.45\linewidth}
\includegraphics[width=\linewidth]{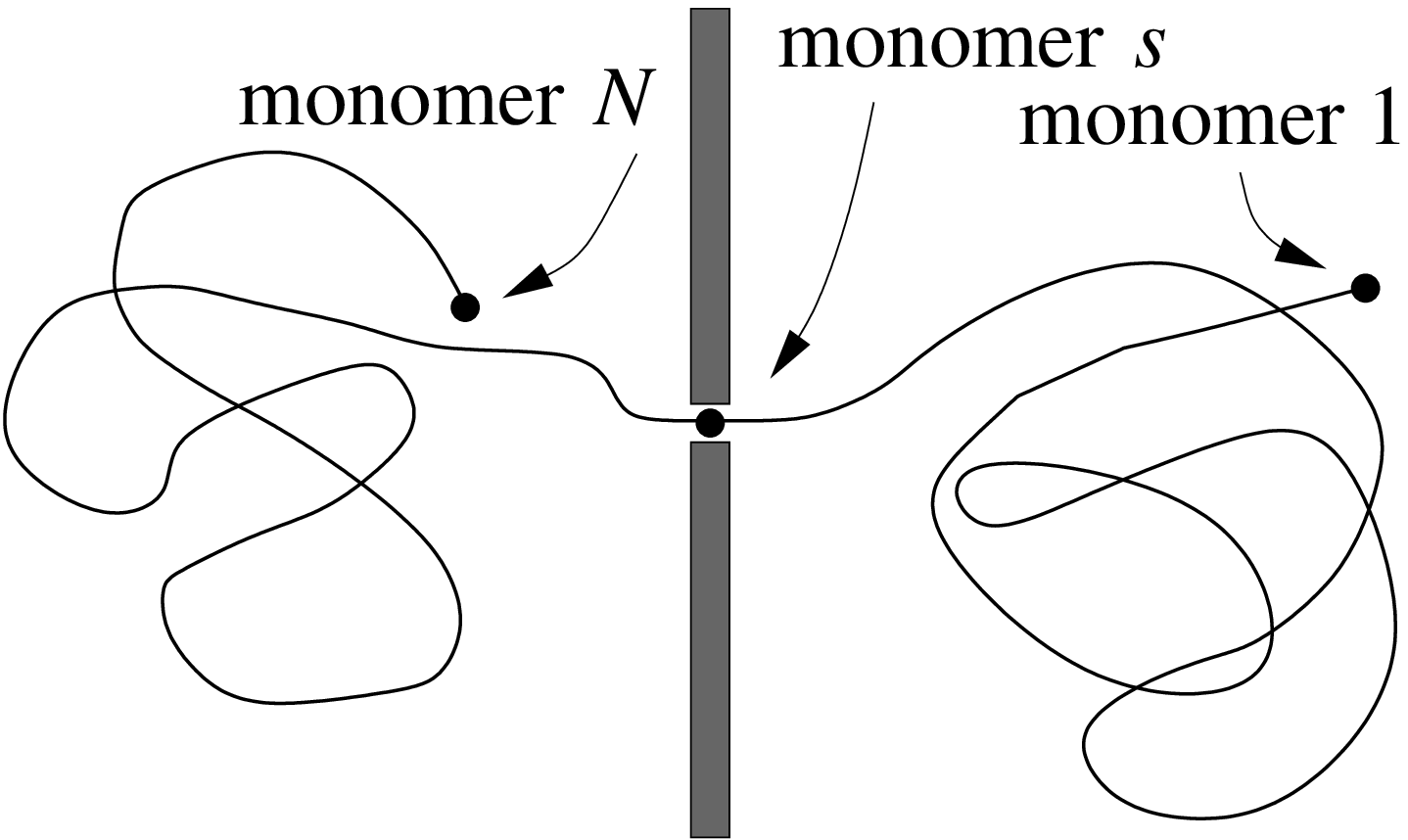}
\end{minipage}
\hspace{0.5cm}
\begin{minipage}{0.5\linewidth}
\includegraphics[angle=270,width=\linewidth]{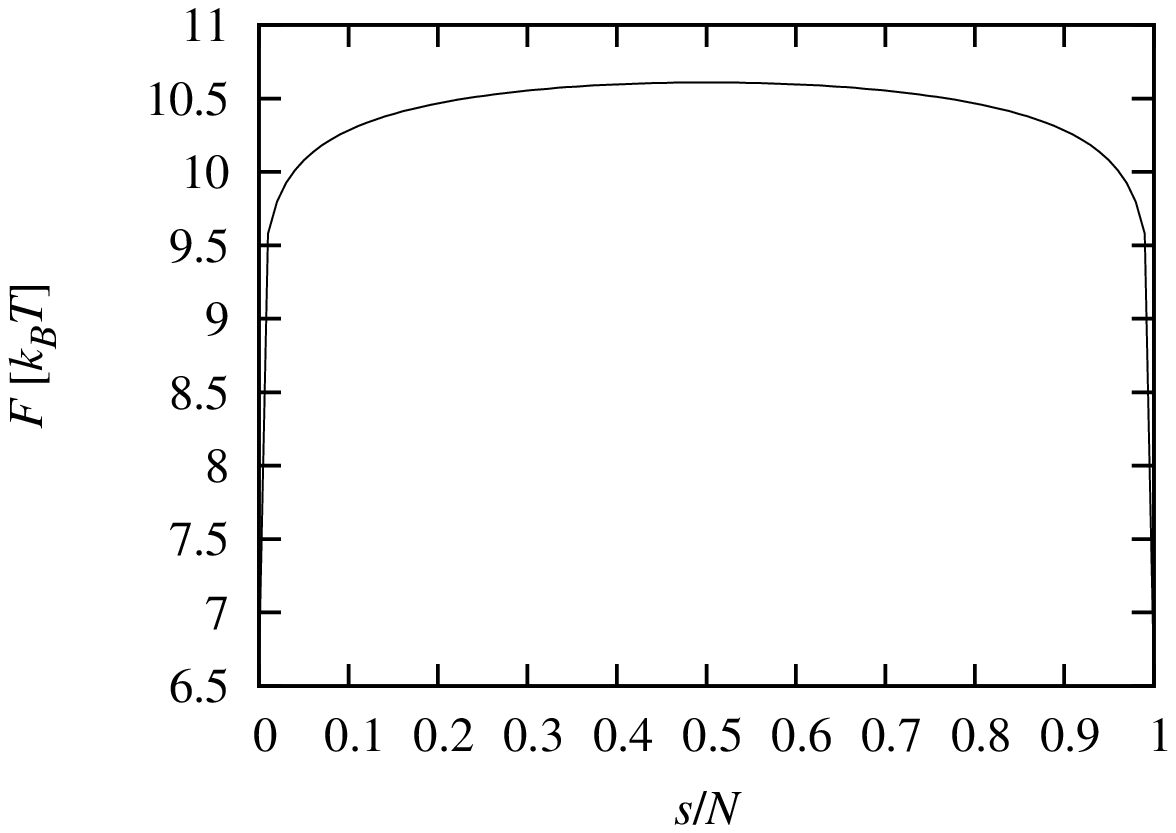}
\end{minipage}
\end{center}
\caption{Left: Snapshot of a translocating polymer in a
  two-dimensional projection. The reaction co-ordinate $s(t)$ denotes
  the monomer index located at the pore at time $t$. Right: The
  entropic barrier as a function of the reaction co-ordinate $s/N$ for
  $N=10^6$. Note that, apart from the first and last $\sim$ 1\%, the
  free energy is within $k_BT$ of its maximal value.}
\label{figst}
\end{figure}

\subsection{Unbiased translocation\label{sec4a}}

To a large extent, the theoretical approach of pore blockade from the
polymer dynamics angle stemmed from the experimental paper by
Kasianowicz {\it et al.} \cite{kasi96}. The early models of a
translocation event reduced polymer dynamics through the pore to the
dynamics of $s(t)$, the index of the monomer located in the pore at
time $t$. For a polymer consisting of $N$ monomers, by definition,
$s(0)=1$ and $s(\tau_d)=N$. Borrowed from chemical physics parlance,
the quantity $s(t)$ (see Fig. \ref{figst}) is termed as the ``reaction
co-ordinate''. The introduction of this quantity allowed the early
researchers to compute the configurational entropy of the chain as a
function of $s(t)$ (concept described in Sec. \ref{sec3a}), rendering
translocation events simply to the motion of an effective single
particle, located at $s(t)$ at time $t$, over an entropic barrier.

\subsubsection{Early works on pore blockade as a quasi-equilibrium
  process\label{sec4a1}}

Aside from the fact that treating translocation as an entropic barrier
crossing process does not filter out the repeated
threading-unthreading sequences that precede the translocation event,
there is another note of caution that needs to be spelled out for this
approach. Since entropy is an equilibrium concept, applying the
entropic barrier concept to study pore-blockade in these early
theories on unbiased translocation \cite{sung96,muthu99} assumes that
at every stage of translocation the polymer has the time to
thermodynamically explore its entire space of configurations, and
therefore the polymer dynamics through the pore is a quasi-equilibrium
process. In any case, from the forms of the partition functions
discussed in Sec. \ref{sec3a}, it is clear that the height of the
entropic barrier as a function of the reaction co-ordinate $s$ is of
the form $\log[s(N-s)]$ (in Refs. \cite{sung96,muthu99} only phantom
polymers were considered, and not self-avoiding polymers, but the form
remains the same), and therefore the barrier is essentially flat
around $s=N/2$ in the scaling limit (polymer length
$N\rightarrow\infty$). The effective particle then has no drive to
move either way on this flat part of the barrier, which stretches for
a length of order-$N$ in the scaling limit, and as a consequence, its
motion is diffusive. The time taken by this effective particle to
cross the barrier --- the pore-blockade time for a translocation event
--- therefore scales as $N^2/D$, where $D$ is the effective diffusion
coefficient of the particle on this entropic landscape.

In Ref. \cite{sung96} the authors assumed that $D$ is a function of
$N$, and used $D\sim N^{-1}$ (resp. $N^{-1/2}$) for phantom Rouse
polymers (resp. phantom Zimm polymers), leading to
$\tau_d\sim N^\alpha$ with $\alpha=3$
(resp. $5/2$). (Strictly speaking, this time is not the true
pore-blockade time as it includes the repeated threading-unthreading
times as well, but in the scaling limit the latter might be
negligible.) Muthukumar \cite{muthu99} subsequently corrected these
results arguing that $D$ should be independent of $N$, which led to
the scaling exponent $\alpha=2$. Slonkina {\it et al.\/} later
generalized these results to channels \cite{slonkina03}.

As it turns out, the quasi-equilibrium approximation is a drastic
simplification as far as scaling results are concerned. We will
discuss its applicability in Sec. \ref{sec4a3}, and discuss its
predictions for field-driven translocation in Sec. \ref{sec4b2}.

\subsubsection{Pore blockade as a non-equilibrium process and anomalous
polymer dynamics\label{sec4a2}}

That the quasi-equilibrium approximation does not hold in the scaling
limit was first pointed out by Chuang {\it et al.} \cite{chuang02}.
They argued that in order to be able to use the quasi-equilibrium
approximation, the polymer needs to have sufficient time to explore
all the accessible configurational states at every value of the
reaction co-ordinate. As the characteristic time for the polymer tails
on the {\it cis\/}- and {\it trans\/}-side increases with polymer
length, the quasi-equilibrium approximation has to break down at some
point. They illustrated this point by considering translocation of a
self-avoiding Rouse polymer: its equilibration time scales as
$N^{1+2\nu}$, which, in the scaling limit will always be far greater
than the scaling of the pore-blockade time $\tau_d\sim
N^2$ predicted by the quasi-equilibrium approximation.

Chuang {\it et al.\/} \cite{chuang02} further argued for a lower limit
for the scaling of the pore-blockade time $\tau_d$ for a
Rouse polymer, as follows. After a translocation event, the polymer
displaces itself by its radius of gyration, which scales as $N^\nu$
for a self-avoiding polymer of length $N$. If the pore width is
infinite (i.e., there is no membrane separating the {\it cis\/} and
the {\it trans\/} sides), then the polymer crosses the pore simply by
diffusion, and the time-scale for crossing the pore follows the
well-known Rouse scaling $N^{1+2\nu}$. When the pore is narrow,
allowing the monomers to pass through only sequentially, the
pore-blockade time can only be larger. They followed up this argument
by computer simulation using the bond-fluctuation model
\cite{carmesin88} (BFM --- a model for self-avoiding Rouse polymer
dynamics), and found that the lower limit $N^{1+2\nu}$ for the scaling
of $\tau_d$ is saturated.  The result meant that the
dynamics during a translocation event is anomalous: if the mean-square
displacement $\langle\Delta s^2(t)\rangle$ of the reaction co-ordinate
in time has to scale as $t^\beta$ for some $\beta$, the condition
$\langle\Delta s^2(\tau_d)=N^2$ along with
$\tau_d\sim N^{1+2\nu}$ means that $\beta=2/(1+2\nu)$
(i.e., $\beta\neq1$) \cite{chuang02}. Their work was quickly ensued by
an incredible body of literature to test these exponents, leading to,
what is colloquially known to researchers in this field, ``an exponent
war''.
\begin{table*}
\begin{center}
\begin{tabular}{c|c|c|c|c}
  Refs. & $\alpha$ (2D, Rouse)&$\alpha$ (2D, Zimm) & $\alpha$ (3D,
  Rouse)& $\alpha$ (3D, Zimm) \tabularnewline \hline \hline  
  \cite{chuang02}&$1+2\nu=2.5$&---&---&---\tabularnewline
  \hline 
&(BFM)&&&\tabularnewline
\cite{luo06a} &$2.50\pm0.01$
  (BFM)&---&---&---\tabularnewline \hline 
  \cite{luo06b} &$2.48\pm0.07$ &---&---&---\tabularnewline  &(FENE
  MD)&&&\tabularnewline \hline \cite{wei07}
  &$2.51\pm0.03$ &---&$2.2$&---\tabularnewline &(bead-spring
  MD)&&$\!$(bead-spring MD)$\!$&\tabularnewline \hline  
  \cite{chatelain08}&2.5 (BFM)&---&---&---\tabularnewline \hline 
  \cite{luo08} &$2.44\pm0.03$ &---&$2.22\pm0.06$ &---\tabularnewline
  & (GROMACS) &---& (GROMACS) &---\tabularnewline \hline 
  \cite{panja07,panja06a}&---&---&$2+\nu\approx2.588$&$1+2\nu\approx2.18$\tabularnewline
  \hline 
  \cite{panja08}&$2+\nu=2.75$&$1+2\nu=2.5$&---&---\tabularnewline \hline
  \cite{dubbeldam07}&---&---&$2.52\pm0.04$&---\tabularnewline  &&&(FENE)
  &\tabularnewline \hline 
  \cite{gauthier08a}&---&---&$2+\nu$ &$1+2\nu$\tabularnewline \hline
  \cite{guillouzic06}&---&---&---&2.27
  (MD)\tabularnewline \hline 
  \cite{gauthier08b}&---&---&---&11/5=2.2 (MD)\tabularnewline \hline
  \cite{kap10}&---&---&---&$2.24\pm0.03$\tabularnewline 
  &&&&(DPD)\tabularnewline \hline
  \cite{haan12}&---&---&$2.516$ (FENE)&\tabularnewline \hline
  \cite{nair12}&---&---&$2.52$ (FENE)&\tabularnewline \hline
\end{tabular}
\caption{Summary of all the results on the exponent for the
  pore-blockade time for unbiased translocation known to us at the
  time of writing this review. Abbreviations are explained in main
  text. \label{table1}}
\end{center}
\end{table*}

For a number of years following the work by Chuang {\it et al.}
\cite{chuang02}, several simulation studies, using BFM, bead-spring
molecular dynamics (MD) and GROMACS, reported the value of $\alpha$
both in 2D and 3D to be consistent with $1+2\nu$ (which equals $2.5$
in 2D, and $\approx2.18$ in 3D \cite{luo06a,luo06b,wei07}. Some of
these studies characterized the anomalous dynamics of unbiased
translocation as well: the mean-square displacement of the monomers
$\langle\Delta s^2(t)\rangle$ through the pore in time $t$ was found
to scale $\sim t^\beta$ with $\beta=2/(1+2\nu)$, satisfying the
obvious requirement $\langle\Delta s^2(\tau_d)\rangle=N^2$. However,
several other subsequent/concurrent theoretical and simulation studies
--- simulations using finitely extensible nonlinear elastic (FENE)
model \cite{fene}, MD and dissipative particle dynamics (DPD) for
modeling hydrodynamic interaction among the monomers \cite{dpd} ---
later found that $\alpha$ and $\beta$ for a Rouse polymer
significantly differed from $1+2\nu$ and $2/(1+2\nu)$ respectively
\cite{wolt06,panja07,panja06a,panja08,dubbeldam07,gauthier08a,guillouzic06,gauthier08b,kap10,haan12,nair12}. The
latter results on the pore-blockade time exponent concentrated around
$2+\nu$ for Rouse, and $1+2\nu$ for Zimm polymers. All these results
are summarized in Table \ref{table1}.

Given that the predictions/confirmations of $\alpha=1+2\nu$ for Rouse
polymer came solely from simulations without a theoretical basis
behind it, while there is a theory that obtains
$\beta=(1+\nu)/(1+2\nu)$, and correspondingly, $\alpha=2+\nu$, we
spend a few sentences on the latter result. It was originally obtained
by two of us \cite{panja07,panja06a,panja08,panja09}, using a
theoretical approach based on polymer's memory effects that stem from
local (in the vicinity of the pore) strain relaxation properties of
polymers, and it was further tested by simulations with a highly
efficient lattice polymer model. The strain results from motions of
monomers across the pore: as a monomer hops from the left to the right
of the pore, the polymer locally stretches on the left and compresses
on the right, giving rise to a local strain, in the form of chain
tension imbalance, across the pore. This imbalance can relax via two
different routes: (i) instantaneously, if the hopped monomer hops
back, and (ii) along the polymer's backbone on both sides of the
membrane, which requires a finite time [the time is simply the Rouse
equilibration time $\tau_R\sim N^{1+2\nu}$, which is the time scale
for the memory of the chain tension to survive]. Until this time the
hopped monomer has an enhanced chance to hop back. When properly
worked out \cite{panja07,panja06a,panja08,panja09}, the memory decays
in time as a power-law, as $t^{-(1+\nu)/(1+2\nu)}$.  This leads to
$\langle\Delta s^2(t)\rangle\sim t^{(1+\nu)/(1+2\nu)}$ [i.e.,
$\beta=(1+\nu)/(1+2\nu)$] up to $\tau_R$, and thereafter as Fickian
diffusion, $\langle\Delta s^2(t)\rangle\sim t$ [i.e.,
$\beta=1$]. These further lead to $\alpha=2+\nu$ ($\approx2.588$ in 3D
and $2.75$ in 2D) \cite{panja07,panja06a,panja08}. For a Zimm polymer,
the memory effects similarly predict $\langle\Delta s^2(t)\rangle\sim
t^{(1+\nu)/(3\nu)}$ up to the Zimm equilibration time $\tau_Z\sim
N^{3\nu}$, and thereafter $\langle\Delta s^2(t)\rangle\sim t$; leading
to the expectation that $\tau_d$ should scale as $N^{1+2\nu}$
\cite{panja07,panja06a}. (The fact that $\alpha=1+2\nu$ for a Zimm
polymer has nothing to do with Rouse dynamics: it is in fact a pure
coincidence that this exponent is the same as that of $\tau_R$
\cite{panja07,panja06a}).

Before we discuss how the apparent differences in the values of
$\alpha$ are reconciled, it would be worthwhile to make a note here on
the attempts to classify the anomalous dynamics of translocation. It
was proposed originally in Ref. \cite{metz03} that anomalous dynamics
of translocation can be expressed in terms of fractional Fokker-Planck
equation, which is based on the continuous time random walk (CTRW)
formalism. Subsequently, some researchers have followed this route
\cite{gros05,dubbeldam07,dubbeldam07a}. The memory effect description
for anomalous dynamics of polymer translocation, on the other hand,
belongs to a general framework of ubiquitous examples of anomalous
dynamics in polymeric systems, based on the generalized Langevin
equation (GLE) \cite{panja10a,panja10b}. The GLE formulation also
establishes that the anomalous dynamics of translocation belongs to
the class of fractional Brownian motion (fBm)
\cite{chatelain08,panja11,dubbeldam11}. Given that fBm and CTRW are
mutually exclusive, the description of polymer translocation using the
fractional Fokker-Planck equation is discredited by these studies.

\subsubsection{Consensus on the value of $\alpha$?\label{sec4a3}}

How can the ``exponent war'' finally end in a truce? Despite
demonstrating that BFM is a pathological model for polymer
translocation \cite{panja10c}, the apparent dispute for the value of
$\alpha$ was left alive and kicking. A recent work by de Haan and
Slater \cite{haan12a} has finally shed an interesting light on this
issue.  They used the FENE model to simulate unbiased polymer
translocation with varying viscosity $\tilde\eta$ of the
surrounding medium. The results are spectacularly consistent with the
memory function approach \cite{panja07,panja06a,panja08,panja09}. At
$\tilde\eta=0$ they found $\alpha=2$ corresponding to Fickian diffusion
for the dynamics of polymer translocation --- this is only to be
expected since the tension imbalance in the vicinity of the pore then
relaxes through the polymer's tails instantaneously, resulting in
complete loss of the memory effects.  As the viscosity is increased,
the apparent exponent $\alpha$ increases, crossing $1+2\nu$, and at
the highest value of viscosity used the apparent exponent $\alpha$
reaches $\approx2.55$ (runs with higher viscosity were not possible
because of prohibitive cost of computation \cite{garycomm}) --- there
is a strong indication from the trend of the data that at very high
viscosity the data would indeed correspond to $\alpha=2+\nu$,
consistent with the prediction of the memory function approach. 

Furthermore, de Haan and Slater \cite{haan12a} showed that if data
from simulations with different viscosities and polymer lengths are
combined, a data collapse can be obtained if $\tau_d/N^2$ is plotted
as a function of $\tilde\eta N^x$ with $x=0.516$. In the limit of high
viscosity, the translocation time is expected to increase linearly
with viscosity. The data collapse at high viscosities predicts
$\tau~/N^2 \sim\tilde\eta N^x$ hence $\alpha=2.516$, close to the
theoretically predicted value $2+\nu$.

\begin{figure}[h]
\begin{center}
\begin{minipage}{0.45\linewidth}
\includegraphics[width=\linewidth]{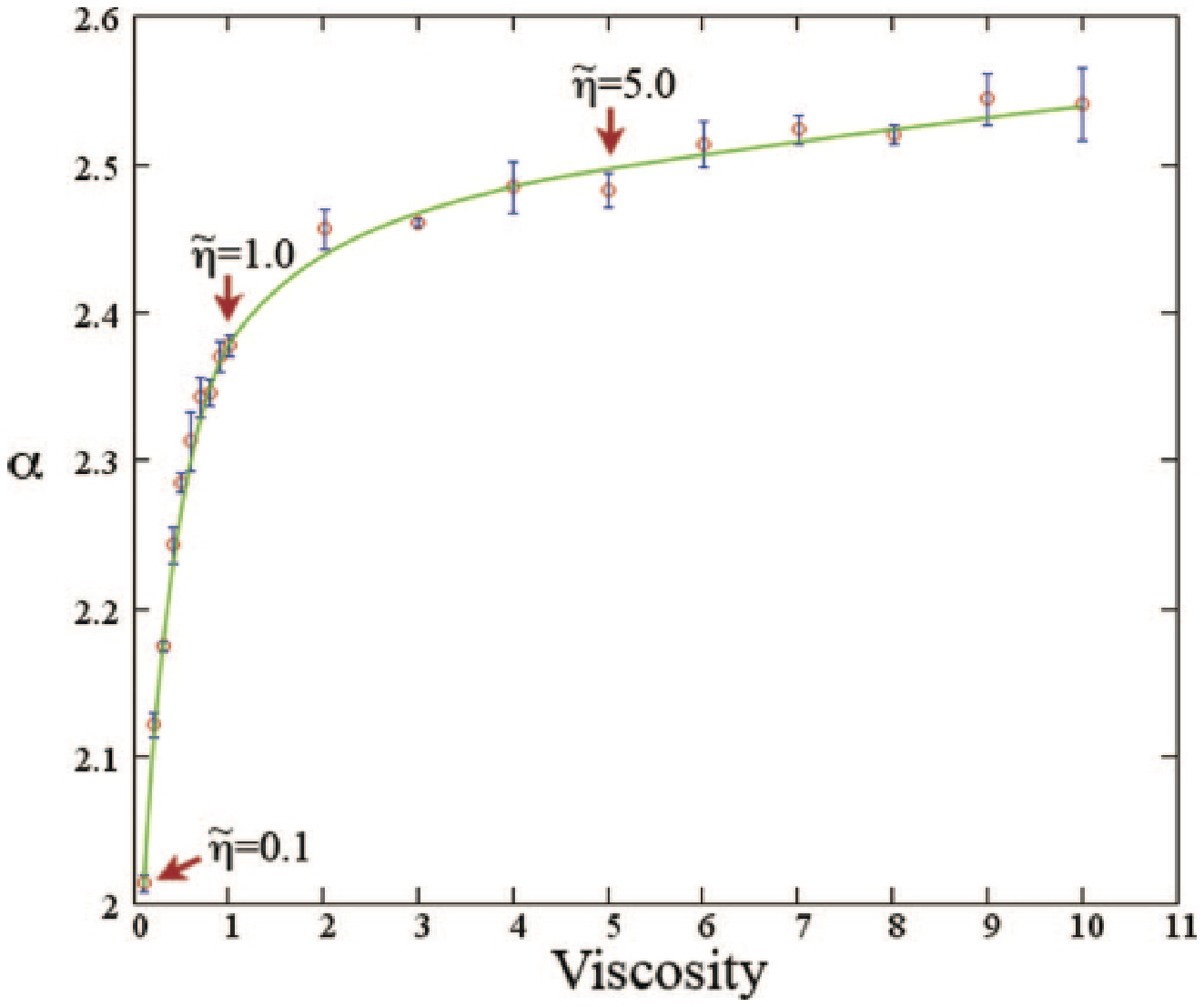}
\end{minipage}
\begin{minipage}{0.45\linewidth}
\includegraphics[width=\linewidth]{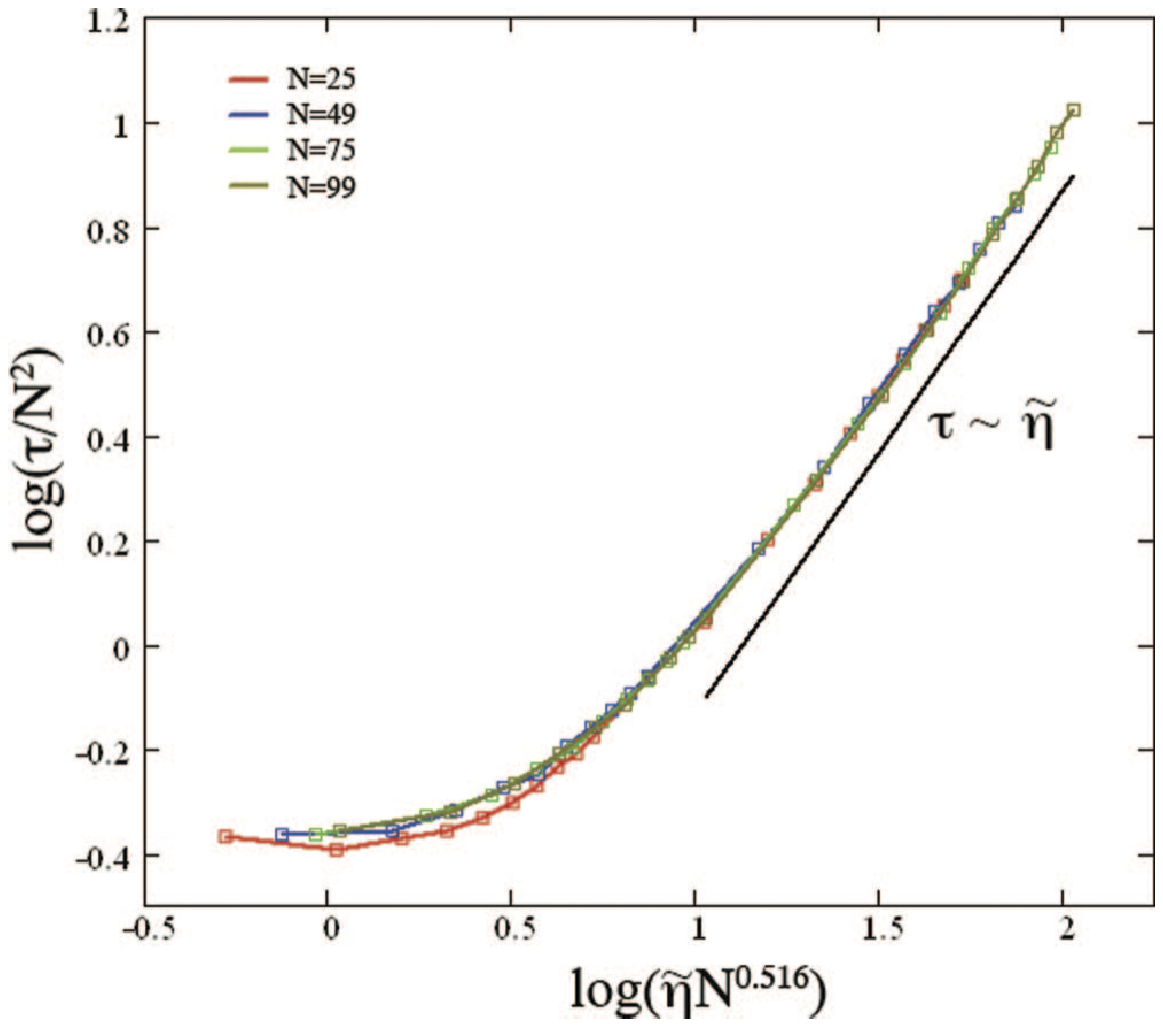}
\end{minipage}
\end{center}
\caption{Results of a FENE simulation model by de Haan and Slater
  \protect\cite{haan12a}. Left: The apparent pore-blockade time
  exponent $\alpha$. Right: Data collapse for $\tau_d/N^2$ as a function
  of $\tilde\eta N^{0.516}$. Reproduced with permission from American
  Institute of Physics.}
\label{garyplot}
\end{figure}

In other words, the work by de Haan and Slater \cite{haan12a} shows
that the true values of $\alpha$ are (i) 2 at zero viscosity and (ii)
likely $2+\nu$ at very high viscosity of the surrounding medium. The
rest of the values reported in the literature are all apparent
exponents.  The corresponding figures by de Haan and Slater is
reproduced in Fig.  \ref{garyplot}. As shown therein, curiously,
choosing $\tilde\eta=1$ in the model produces (the apparent exponent)
$\alpha=1+2\nu$ (the other arrow at $\tilde\eta=5$ indicates the
results of another study by the authors).

\subsection{(Electric) field-driven translocation\label{sec4b}}

The extension from unbiased translocation to (electric) field-driven
translocation is in principle trivial: one simply adds a force on the
monomers equaling the charge of the monomers (the charge per monomer
is henceforth understood to be unity without any loss of generality)
times the strength of the electric field $E$, acting from the {\it
  cis\/} to the {\it trans\/} direction. As explained in
Sec. \ref{sec3}, this is a reasonably accurate approximation since the
electric field dies off rapidly within the Debye length, which is less
than a nanometer under typical experimental conditions. However, as we
will soon see, the presence of the electric field complicates the
scaling issues.

\subsubsection{Extension of the quasi-equilibrium
  picture\label{sec4b1}}

\begin{figure}[h]
\centering
\includegraphics[angle=270,width=0.55\linewidth]{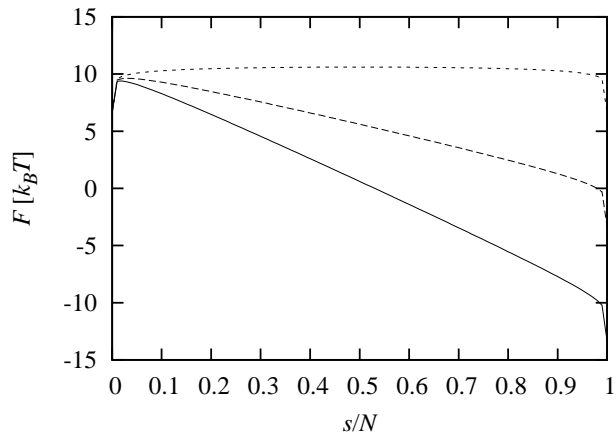}
\caption{The tilted entropic barrier as a function of the reaction
  co-ordinate $s/N$ in the presence of a small driving field for
  $N=10^6$ (the values used for the field are 0.5 and 0.25 $\mu$eV per
  monomer respectively: higher tilt corresponds to stronger
  field). The no-tilt case is shown as a reference in small dashed
  line when there is no field acting in the pore (i.e., unbiased
  translocation). The tilt gets stronger with increasing field
  strength.}
\label{fieldtilt}
\end{figure}

When such a field is added to the equation of motion in the
quasi-equilibrium description of translocation, the entropic barrier,
in terms of the reaction co-ordinate $s$, gets an overall linear tilt
from the {\it cis\/} towards the {\it trans\/} side
(Fig. \ref{fieldtilt}). The result of this exercise is that on top of
the diffusive motion as described in Sec. \ref{sec4a1}, the effective
particle also has a constant drift towards the {\it trans\/} side,
meaning that it traverses the entire length $N$ of the entropic
barrier with a uniform velocity, which is proportional to the field
strength. Consequently, the pore-blockade time simply scales as $N$
and is inversely proportional to the field strength
\cite{muthu99,lub99,slonkina03}.

It is to be noted that for the above to work that {\it per se\/} one
does not need to have an electric field acting on the monomer
straddling the pore. The field can have multiple origin, such as
entropic due to (preferential) confinement
\cite{wong08,park98,muthu01} or adsorption to the membrane on the {\it
  trans\/} side \cite{park98a}; in these cases the field $E$ acting on
each monomer in the pore for field-driven translocation is simply
replaced by the chemical potential gradient $\Delta\mu$ for monomer
transfer from the {\it cis\/} to the {\it trans\/} side, and we will
summarily refer to all these situations as field-driven translocation.

\subsubsection{Pore blockade as a non-equilibrium process\label{sec4b2}}

Just like in the case of unbiased translocation, that the
quasi-equilibrium picture cannot hold in the scaling limit was first
pointed out by Kantor and Kardar \cite{kantor04}. They considered the
case of a self-avoiding Rouse polymer driven by a field acting within
the pore and argued that the pore-blockade time has a lower limit which scales
as $N^{1+\nu}$, as follows. Consider a pore of infinite diameter,
i.e., the motion of a free polymer of length $N$, of which one of the monomers
is being pulled by a force $E$. The motion of the center of
mass of this polymer determined by the total force acting on the polymer,
and thus results in a uniform velocity $\propto E/N$. When
the polymer completely translocates, it displaces its center of mass
by a distance that scales as its radius of gyration $N^\nu$, i.e., the
total time of translocation then scales as $N^{1+\nu}/E$. When the
pore is narrow --- for which the field acts on the monomer
instantaneously located within the pore, but that is a matter of
detail --- the pore-blockade time cannot be less than the case when
the pore is infinitely wide, leading to the lower limit for the
pore-blockade time for translocation as $N^{1+\nu}/E$. They carried
out simulations with the BFM in two dimensions, and concluded that this
lower limit is saturated, i.e., $\langle\tau_d\rangle\sim N^{1+\nu}/E$.

Again, this study was ensued by a rather large number of follow-up
ones to confirm the result, and led to another exponent war, only
messier. As we shall shortly see, this has to do with the fact that
the addition of a driving field into the problem introduces an extra level of
complication; however, the reported values of the exponent all fall in
a consistent line once the scaling limit $(N\rightarrow\infty)$ is
properly interpreted in relation to $E$.
\begin{figure}[h]
\centering
\includegraphics[width=0.25\linewidth]{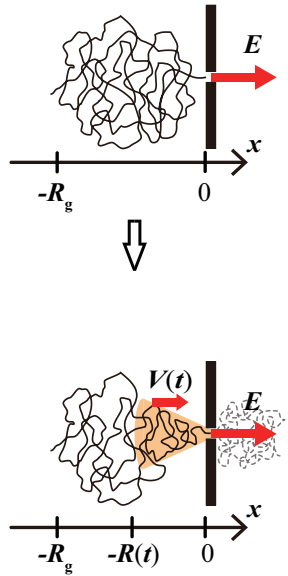}
\caption{Schematic figure illustrating the ``tension propagation
  theory'' by Sakaue and coworkers. We thank T. Saito for providing
  us with this figure.}
\label{sakauefig}
\end{figure}

Rather than providing a chronological narrative for the values of
$\alpha$ as reported by different research groups, we opt to first
present the theoretical perspective as followed by Sakaue and
coworkers, as we feel that this is the most robust description. A
schematic representation of this process is shown in
Fig. \ref{sakauefig}: for $E>k_BT/R_g$ they identified that the
polymer on the {\it cis\/} side is composed of two domains, a moving
one [moving with velocity $V(t)$; the range of this domain extends up
to a distance $R(t)$ from the pore on the {\em cis\/} side, as shown
in Fig. \ref{sakauefig} with a colored background], and a quiescent
one [beyond distance $R(t)$ from the pore]; and the key to
field-driven translocation dynamics is the shifting boundary between
the two domains of the polymer, located at a distance $R(t)$ from the
pore on the {\it cis\/} side, as it determines how the driving field
is transmitted along the backbone
\cite{sakaue07,sakaue10,sakaue11,sakaue12a,sakaue12b}. The moving
domain corresponds to a velocity and force-extension relation dictated
by the those of a polymer in the ``trumpet regime'', while the
quiescent domain corresponds to those of a polymer essentially
unperturbed by the applied field. Matching the boundary conditions
between the two domains then leads to the behavior of $R(t)$ as a
function of $t$, and subsequently the pore-blockade time is determined
from the relation $R(\tau_d)=R_g\sim N^\nu$. In other words, the
pore-blockade time is dominated by the tension propagation time along
the backbone of the chain. In this way $\alpha$ is shown to be equal
to $1+\nu$, although the exponent for the field-dependence of the
pore-blockade time depends on the field strength, i.e., whether
$k_BT/R_g<E<k_BT/a$ or $E>k_BT/a$, where $a$ is the length of a
monomer.

The same exponent has been obtained theoretically by Rowghanian and
Grosberg \cite{rowgh11} and Dubbeldam {\it et al\/}
\cite{dubbeldam12}, using ``iso-flux trumpet'' models. These models
entail small variations of those of Sakaue and coworkers, and posit
that instead of the polymer attaining the shape of a trumpet on the
{\it cis\/} side, one should imagine a space-fixed trumpet, expanding
in radius away from the pore on the {\it cis\/} side, and the polymer
has to funnel through this trumpet. At any given time, the flux of
monomers is uniform across any cross-section of the trumpet.

All these models presuppose that under the influence of the field the
polymer takes a far-out-of-equilibrium shape; this assumption
obviously has to break down if the field becomes small enough. At
small enough fields the translocation dynamics can be simply extended
as a linear response on the unbiased case, for which the memory
function approach has done well, by adding a force on the monomer
straddling the pore. The result is that in three dimensions, the
pore-blockade time exponent $\alpha$ for weak fields is obtained from
the exponent for the memory function, namely that
$\alpha=(1+2\nu)/(1+\nu)$ for a Rouse polymer, and $=3\nu/(1+\nu)$ for
a Zimm polymer \cite{vocks08}.

How can one distinguish weak and strong forces from each other?  One
assumption underlying the memory function approach applied to
field-driven translocation is that the polymer's configurational
statistics is not influenced by the applied field. A
back-of-the-envelope calculation, analogous to the well-known coil to
Pincus blob transition in polymer physics, leads to the result that
the field strength satisfying the condition $E^*R_g=k_BT$ decides
whether the field is weak or strong. This relation has recently been
verified by Sakaue for the dynamics of a free polymer in bulk
\cite{sakaue_unpub}. With $R_g\sim N^\nu$, this condition entails the
following scenario: (i) for a Rouse polymer in two dimensions the
pore-blockade time exponent is $2\nu=1.5$ and $1+\nu=1.75$ for weak
$(E<E^*)$ and strong $(E>E^*)$ fields respectively, and (ii) for a
Rouse polymer in three dimensions the pore-blockade time exponent is
$(1+2\nu)/(1+\nu)\approx1.37$ and $1+\nu\approx1.588$ for weak
$(E<E^*)$ and strong $(E>E^*)$ fields respectively.  Note that the
crossover field strength $E^*$ is decreasing with polymer
length. Thus, no matter how small the field is, if the polymer gets
long enough, the behavior will eventually cross over from the linear
response regime to the regime described by Sakaue et al. This
crossover was indeed argued (and shown numerically) by us in a related
paper that mapped the dynamics of polymer adsorption to that of
polymer translocation \cite{panja09a}, wherein the adsorbing force
(derived from the adsorbing energy) plays the role of the
translocating force generated by the field. The argument rests on the
assumption that at strong fields the polymer attains the so-called
``stem-flower'' configuration \cite{broch95}, with a quiescent
``flower'' consisted of the monomer cloud connected by a ``stem'' of
monomers to the adsorbing surface; the flux of monomers are brought to
the adsorbing surface along the stem by the action of the adsorbing
force. 

For one, it is clear from the above discussions that the reported
value of the pore-blockade time exponent has the potential to be
easily influenced by the model parameters: in particular, how the
scaling limit $(N\rightarrow\infty)$ is interpreted in comparison to
$E$ (whether it is bigger or smaller than $E^*\sim N^{-\nu}$). In our
view, this is the reason why establishing these exponents in
simulations has been no trivial matter. A second source of
complication is that Brownian and Langevin dynamics simulations very
often introduce a pore friction, which also considerably influences
the measured value of the exponent. Nevertheless, in Tables
\ref{table2} and \ref{table3} we report that there is quite some
numerical support from the different research groups for $\alpha=2\nu$
and $\alpha=1+\nu$ for a Rouse polymer in two dimensions, and
$\alpha=(1+2\nu)/(1+\nu)$ and $\alpha=1+\nu$ for a Rouse polymer in
three dimensions.\footnote{The list in Tables \ref{table2} and
\ref{table3} is by no means exhaustive. There are papers, such as
Refs. \cite{lehtola08,lehtola09,lehtola10,metz09,metz10}, who report
extensively on the effective pore-blockade time exponent as
functions of simulation parameters; we have found them difficult to
interpret for including them in this review in a cogent
manner.\label{footfieldtrans}}
\begin{table}[h]
\begin{center}
\begin{tabular}{c|c}
  Refs. & $\alpha$ (2D, Rouse) \tabularnewline \hline \hline  
  \cite{kantor04}&$1+\nu=1.75$ (BFM)\tabularnewline \hline 
  \cite{luo06c}& $1.46\pm0.01$ crossing over to $1.73\pm0.02$ with
  increasing $N$ at fixed $E$ (BFM)\tabularnewline \hline 
  \cite{huo06} &$1.50\pm0.01$ crossing over to $1.69\pm0.04$ with
  increasing $N$ at fixed $E$ (LD)\tabularnewline \hline 
  \cite{cac06}  &$1.55\pm0.04$ (MC) \tabularnewline \hline  
  \cite{panja08} &$2\nu=1.5$\tabularnewline\hline 
\end{tabular}
\caption{Summary of all the reported values of the exponent for the
  pore-blockade time for field-driven translocation of a Rouse polymer
  in two dimensions, known to us at the time of writing this
  review. Abbreviations: LD (Langevin dynamics), MC (Monte
  Carlo). \label{table2}}
\end{center}
\end{table}
\begin{table}[h]
\begin{center}
\begin{tabular}{c|c}
  Refs. & $\alpha$ (3D, Rouse) \tabularnewline \hline \hline  
  \cite{wei07}&1.27\tabularnewline \hline 
  \cite{milchev04}&$1.65\pm0.08$\tabularnewline\hline 
  \cite{luo08a}&$1.42\pm0.01$ (MD, LD)\tabularnewline \hline 
  \cite{dubbeldam07a}&1.5 (FENE)\tabularnewline \hline
  \cite{vocks08}&$\displaystyle{\frac{(1+2\nu)}{1+\nu}\approx1.37}$\tabularnewline \hline
  \cite{bhatt09}&$1.36\pm0.01$ \tabularnewline \hline
  \cite{fyta08}&$1.36\pm0.03$  (MD)\tabularnewline \hline 
\cite{nair12}&$1.35$-$1.40$ (LD) \tabularnewline \hline
\end{tabular}
\caption{Summary of all the reported values of the exponent for the
  pore-blockade time for field-driven translocation of a Rouse polymer
  in three dimensions, known to us at the time of writing this
  review. Abbreviations: LD (Langevin dynamics), MC (Monte
  Carlo). \label{table3}}
\end{center}
\end{table}

Although in our view the results for two dimensions can clearly be
reconciled by energy conservation arguments \cite{panja08}, and the
memory function approach at weak fields ($E<E^*$) and
stem-flower/tension propagation/trumpet models at moderate to strong
fields ($E>E^*$) in three dimensions, this is by no means the only
interpretation. In two recent papers Ikonen et
al. \cite{iko12a,iko12b} reanalyzed some of their own older data as
well as those of Lehtola {\it et al.\/} \cite{lehtola08}; using pore
friction as a control variable in a Brownian dynamics tension
propagation scheme they collapsed all the data on a master curve to
establish that all data points to the pore-blockade time
scaling as $N^{1+\nu}$.\footnote{Further, the scatter in the
  dependence of pore-blockade time on $E$, reported in several
  simulation papers \cite{dubbeldam12,lehtola08,metz09,metz10,iko12b}
  is far too big to draw a definitive conclusion: aside from the
  complications involving model parameters as noted above, it is not
  easy to let the field value span multiple decades such that a
  power-law dependence can be determined reliably; although there is a
  predominance of reporting $\tau_d\sim1/E$.}

One last, theoretically interesting remark:
note that the value $\alpha=(1+2\nu)/(1+\nu)$ for a Rouse polymer in
three dimensions violates the expected lower limit $1+\nu$ expected by
Kantor and Kardar \cite{kantor04}. This is because the lower limit
proposed by them is incorrect. This has been easily argued from energy
conservation considerations \cite{panja09a,panja08}. Consider a
translocating Rouse polymer under a field $E$: $N$ monomers take time
$\tau_d$ to translocate through the pore. The total work done by the
field in time $\tau_d$ is then given by $EN$. In time $\tau_d$, each
monomer travels a distance $\sim R_g\sim N^\nu$, the radius of
gyration of the polymer, leading to an average monomer velocity
$v_m\sim R_g/\tau_d$. The rate of loss of energy due to viscosity
$\tilde\eta$ of the surrounding medium per monomer is given by
$\tilde\eta v_m^2$. For a Rouse polymer, the frictional force on the
entire polymer is a sum of frictional forces on individual monomers,
leading to the total free energy loss due to the viscosity of the
surrounding medium during the entire translocation event scaling as
$\Delta F\sim N\tau_d\tilde\eta v_m^2 =N\tilde\eta R_g^2/\tau_d$. This
loss of energy must be less than or equal to the total work done by
the field $EN$, which yields us the inequality $\tau_d\ge\tilde\eta
R_g^2/E = \tilde\eta N^{2\nu}/E$. In three dimensions
$2\nu<(1+2\nu)/(1+\nu)$, so the result that $\alpha=(1+2\nu)/(1+\nu)$
for a Rouse polymer does not violate the lower limit [a similar
argument leads to the result result $\alpha=3\nu/(1+\nu)$ for a Zimm
polymer also does not violate the corresponding lower limit $3\nu-1$].

In contrast to the above, field-driven translocation of Zimm polymers
has been studied with much less intensity. The memory function
approach predicts $\alpha=3\nu/(1+\nu)\approx1.11$ that should only
hold for weak fields \cite{vocks08}. Unlike Rouse polymers,
field-driven translocation of Zimm polymers is accessible by
experiments \cite{storm05b}, reporting a pore-blockade time exponent
$1.27\pm0.03$. The authors explained the exponent to be $2\nu$, based
on a mechanistic picture wherein the polymer chain on the {\it cis\/}
side moves as a macroscopic blob, as it gradually gets sucked into the
pore. While such a mechanistic picture is unlikely to be correct, we
note that simulation results confirming the numerical value of the
exponent does exist ($\alpha=1.28\pm0.01$) \cite{fyta08}. There is
however another simulation study reporting $\alpha\approx1.2$,
claiming an agreement with $\alpha=3\nu/(1+\nu)$, the prediction from
the memory function approach. The existence of $E^*$ as distinguishing
strong and weak field regimes, and the dependence of the pore-blockade
time on the strength of the applied field have essentially not been
addressed.

\subsection{Translocation by pulling with  optical
  tweezers\label{sec4c}}

In this method of translocation a fluorescent bead is attached to one
end of the polymer after the polymer is threaded through the pore. The
bead is then captured by optical tweezers and as it is pulled away
from the pore on the {\it trans\/} side, the rest of the polymer
translocates through the pore. This experiment is motivated by the
desire to determine the secondary structure of a RNA molecule (see
Ref. \cite{vocks09} and the references cited therein).

Several groups have studied this problem for Rouse polymers, and
$\tau_d\sim N^2$ has been unambiguously established
\cite{huo07,panja08a} --- this is a rare case of agreement in this
field. The memory function approach predicts this exponent
\cite{vocks09}: on the {\it trans\/} side of the membrane the polymer
achieves a stretched configuration, leading to a power-law memory in
time that behaves as $t^{-1/2}$, while on the {\it cis\/} side the
polymer's memory decays in time, for weak pulling force, as
$t^{-(1+2\nu)/(1+\nu)}$. The first one, being the slower of the two,
determines the pore-blockade time exponent. This picture is also
confirmed by simulations of polymer translocation under a double force
arrangement \cite{ollila09}, and by pulling an adsorbed polymer away
by an optical tweezers \cite{paturej12}, a problem that can be mapped
to translocation by pulling force just like the adsorption problem has
been mapped on to field-driven translocation problem \cite{panja09a}.

\subsection{Epilogue to Sec. \ref{sec4}\label{sec4d}}

The reader should bear in mind that the purpose behind Sec. \ref{sec4} is {\it not\/} to provide an encyclopedic summary of all the
translocation studies on pore-blockade times; instead, the purpose is
to present the generic problems that several research groups have
concentrated on. There are many studies like polymer translocation
through pores with complex geometries, such as
Refs. \cite{mohan10,wong07} as well as pioneering theoretical studies on protein translocation across nanopores using Langevin dynamics and molecular dynamics simulations (Refs. \cite{makarov08,tian05,huang05,huang08a,huang08b,kirmizialtin04,kirmizialtin06}) that we have chosen to leave out. In this
context we note that there are some interesting problems like
zipping-unzipping dynamics of DNA strands that have been mapped on to
translocation \cite{ferr10,ferr11}. To what extent these are related
to translocation is, however, unclear.

\section{Experimental aspects still in the want of theoretical understanding\label{sec5}} 

The number and variety of experimental investigations that target
polymer dynamics aspects during the translocation is constantly
increasing. Many new techniques have recently been developed for
uncovering mechanisms of transport through channels and nanopores,
providing an increased amount of microscopic information on
translocation processes. Some of these experimental studies have been
addressed by theoretical works, but there is still a significant
number of observations that need to be fully explained.  In this
section our goal is to highlight some of these polymer translocation
phenomena that are still not well-understood.

The success of current nanopore translocation methods is based on very
precise measurements of current fluctuations of small ions during the
experiments
\cite{zwolak08,meller03,muthu_book,keyser11,aksimentiev11}. These
fluctuations appear because fluxes of charged particles present in the
system across the channel are different with and without the polymer
in the pore. The majority of experiments have reported a drop in the
current (a current blockade) when the polymer threads through the pore
\cite{zwolak08,meller03}. It has been argued that these observations
can be well-understood since the polymer geometrically excludes some
part of the channel from small cations and anions, reducing the
overall flux of charged particles. However, a recent experiment on
translocation of double-stranded DNA molecules through some
solid-state nanopores reported a surprising {\it increase} in the
channel current, observed at some sets of parameters \cite{smeets06},
and these results have challenged existing views on nanopore sensing
as a purely geometric exclusion phenomenon. It was shown that at low
concentrations of the salt (specifically, KCl in experiments) the
current during the translocation increases, while for large
concentrations there are current blockades. Several phenomenological
models of how the charged particle flux through nanopores can be
enhanced have been proposed \cite{smeets06}; in particular, it was
argued that the presence of a negatively charged DNA molecule attracts
additional $K^{+}$ cations into the channel, and the overall current
might increase. However, these approaches produced simplified and very
qualitative descriptions that have led to many contradictions and
questions. For example, if DNA attracts cations into the pore why do
these ions leave the channel?  The strong attraction into the pore
would lead to lowering the current, in contrast to observations in
these experiments. Why does the attraction exist only for some sets of
parameters, e.g., low-salt conditions?  Why are these phenomena
observed only for artificial solid-state nanopores, while dynamics in
biological channels is more or less consistent with exclusion
arguments?  From these questions one can conclude that the microscopic
origins of this unusual phenomenon are still far from being
clear. This is a critical issue since all quantitative data on polymer
translocation are associated with changes in currents of charged
particles, which apparently are not understood.

Another important problem that needs to be resolved theoretically is
connected with the role of polymer conformations during
translocation. It is widely assumed that during the motion through the
channel the polymer molecule moves as a linear chain in a
``single-file'' fashion. However, experiments on solid-state nanopores
suggest that in many cases the polymer translocates in the partially
folded conformation \cite{storm05a,storm05b}. In this case, the
polymer experiences spatially and temporary varying interactions with
the pore, leading to complex dynamics. Similar problems are observed
in experiments where the polymer adsorbs near the nanopore, modifying
the current through the channel \cite{vlassarev12}. The translocation
of folded polymers has been addressed theoretically, but only using
the quasi-equilibrium phenomenological approach \cite{kotsev07}, while
in this case one expects non-equilibrium phenomena to have a
stronger influence on polymer dynamics. The important questions here
are the following: (i) Are polymer folding conformations affecting the
translocation dynamics? (ii) What is the role of polymer-pore
interactions in this case? (iii) How is the channel geometry coupled
to translocation of the folded polymers? It is important to develop a
non-equilibrium approach that will address these important issues.

\section{Perspective: the future of this field\label{sec6}}

In recent years, the field of polymer translocation has seen a fast
growth with strong advances. It is now possible to observe
single-molecule polymer dynamics during the motion through channels
with unprecedented spatial and temporal resolution. These striking
experimental studies stimulated many theoretical
developments. However, although several ideas that underlie the
non-equilibrium nature of polymer translocation have been introduced
and tested in extensive theoretical and computer simulation studies,
most experiments are still analyzed using over-simplified
quasi-equilibrium theoretical methods. It is important to extend the
non-equilibrium approaches to describe not only computer simulations
but more importantly real polymer translocation phenomena (as in
experiments). In our opinion, this will be one of the most difficult
challenges for the field.

Considering theoretical advances in the translocation, one can see
that several mechanisms to understand the deviations from equilibrium
dynamics of threading polymers have been proposed and
analyzed. Because of these developments many features of polymer
transport through channels are now better understood. However, none of
existing methods can fully explain polymer dynamics in all parts of
the parameters space. It suggests that there is a need to develop a
unified comprehensive theoretical approach that will fully address all
issues associated with the non-equilibrium nature of polymer
translocation and that will be valid for all conditions.  This will be
another important goal for future theoretical studies on
translocation. It is clear that future progress on understanding the
mechanisms of polymer motion through channels and pores will strongly
depend on combined theoretical and experimental efforts to analyze
these complex phenomena.

\section*{Acknowledgement}

We thank Takahiro Sakaue for detailed comments on the manuscript, and
the wider translocation community for many lively discussions. A.B.K. acknowledges the support from the Welch Foundation (Grant C-1559).

\end{document}